\documentclass[twocolumn,prb,superscriptaddress,longbibliography,showpacs]{revtex4-1}
\usepackage{revsymb4-1}
\usepackage{array}
\usepackage{amsfonts}
\usepackage{amssymb}
\usepackage[dvips]{graphicx}
\usepackage{color}
\usepackage{amsmath}

\def\tablename{Table}
\def\figurename{Figure}
\makeatletter
\renewcommand{\fnum@figure}[1]{\textbf{\textsf{\figurename~\thefigure}} $\mathbf{|}$ }
\renewcommand{\fnum@table}[1]{\textbf{\textsf{\tablename~\thetable}} $\mathbf{|}$ }
\makeatother

\makeatletter
\providecommand{\tabularnewline}{\\}

\@ifundefined{textcolor}{}
{%
 \definecolor{BLACK}{gray}{0}
 \definecolor{WHITE}{gray}{1}
 \definecolor{RED}{rgb}{1,0,0}
 \definecolor{GREEN}{rgb}{0,1,0}
 \definecolor{BLUE}{rgb}{0,0,1}
 \definecolor{CYAN}{cmyk}{1,0,0,0}
 \definecolor{MAGENTA}{cmyk}{0,1,0,0}
 \definecolor{YELLOW}{cmyk}{0,0,1,0}
 }
\makeatother
\newcommand{\bra}[1]{\langle #1|}
\newcommand{\ket}[1]{|#1\rangle}
\newcommand{\braket}[2]{\langle #1|#2\rangle}
\renewcommand{\t}[1]{\textrm{#1}}
\newcommand{\mat}[2]{
\left(\begin{array}{#1}
#2
\end{array}
\right)}

\begin{document}

\title{The elusive Heisenberg limit in quantum enhanced metrology}

\author{Rafa\l{} Demkowicz-Dobrza\'{n}ski}
\affiliation{University of Warsaw, Faculty of Physics, ul. Ho\.{z}a 69, PL-00-681 Warszawa, Poland}
\author{Jan Ko\l{}ody\'{n}ski}
\affiliation{University of Warsaw, Faculty of Physics, ul. Ho\.{z}a 69, PL-00-681 Warszawa, Poland}
\author{M\u{a}d\u{a}lin Gu\c{t}\u{a}}
\affiliation{University of Nottingham, School of Mathematical Sciences, University Park, NG7 2RD Nottingham, UK}

\begin{abstract}
Quantum precision enhancement is of fundamental importance for the
development of advanced metrological optical experiments such as gravitational wave detection
and frequency calibration with atomic clocks.
Precision in these experiments is strongly limited by the $1/\sqrt{N}$
shot noise factor with $N$ being the number of probes (photons, atoms) employed in the experiment.
Quantum theory provides tools to overcome the bound by using entangled probes.
While in an idealized scenario this gives rise to the \emph{Heisenberg scaling} of precision  $1/N$,
we show that when decoherence is taken into account, the maximal possible quantum enhancement
amounts generically to a constant factor rather than quadratic improvement. We provide efficient
and intuitive tools for deriving the bounds
based on the geometry of quantum channels and semi-definite programming.
We apply these tools to derive bounds for models of decoherence relevant for metrological applications including: dephasing,
depolarization, spontaneous emission and photon loss.

\end{abstract}


\maketitle

Quantum enhanced metrology aims to exploit quantum features of atoms and light
such as entanglement, for measuring physical quantities with precision going beyond the classical limit
\cite{Giovannetti2011, Maccone2011, Paris2009, Banaszek2009}.
A prominent example is that of an optical interferometer, where
interference of photons at the output port carries information on the relative optical path difference between the
interferometer arms. When standard laser light is used, the observed results are compatible with the claim that
 ``each photon interferes only with itself'' \cite{Dirac1958} and the whole process may be regarded as
 sensing with $N$ independent probes (photons).
Parameter estimation with $N$ independent probes yields the $1/\sqrt{N}$ \emph{standard scaling} (SS) of precision
\cite{Guta2009}.
Entangling the probes however, can in principle offer a quadratic enhancement in precision, i.e. the $1/N$ or \emph{Heisenberg scaling} (HS)
\cite{Giovannetti2004,Giovannetti2006,Braunstein1994,Berry2000,Zwierz2010}.
Such strategies have been experimentally realized in optical interferometry \cite{Mitchell2004,Eisenberg2005,Nagata2007,Resch2007,Higgins2009} with exciting applications in the quest for the first direct detection of gravitational waves \cite{LIGO2011,Goda2008}.
Moreover, the same quantum enhancement principle can be utilized in atomic spectroscopy \cite{Wineland1992,Huelga1997} where the spin-squeezed states have been employed for improving frequency calibration precision
 \cite{Meyer2001,Leibfried2004,Wasilewski2010,Koschorreck2011}.
Alternative approaches to beat the SS without resorting to quantum entanglement include
multiple-pass \cite{Higgins2007, Demkowicz2010} and non-linear metrology \cite{Boixo2007, Napolitano2011}.

Unfortunately, both the theory \cite{Dorner2008,Demkowicz2009a,Shaji2007,Andre2004,Meiser2008} and experiments \cite{Kacprowicz2010}
 confirmed the fragility of the above schemes when noise sources such as \emph{decoherence} are considered, and
 it has been rigorously shown for particular models that asymptotically with respect to $N$, even infinitesimally
  small noise turns HS into SS, so that the quantum gain amounts to a constant factor improvement \cite{Kolodynski2010,Knysh2011, Escher2011}.

In this paper we develop two methods which allow to extend these partial results to a broad
class of decoherence models, and in particular to obtain fundamental bounds on quantum enhancement
for the most relevant models encountered in the quantum metrology literature.
The two methods complement each other in terms of provided intuition and power.

The first one elaborates on the idea of classical simulation (CS) \cite{Matsumoto2010}
and provides a bound based solely on the geometry of the space of quantum channels (SQC).
When applicable, it gives an excellent intuition as to why the HS is lost in the presence of decoherence, but fails to yield useful bounds for some relevant decoherence models.

The second one is based on the channel extension (CE) method \cite{Fujiwara2008} and
requires the optimization over different Kraus representation of a quantum channel. However,
unlike the earlier work\cite{Escher2011} which involved making an educated guess about the appropriate class of Kraus representations, our bound can be cast into an explicit semi-definite optimization problem which is easy to solve even for complex models.The power of the method is demonstrated by obtaining new bounds for depolarization and spontaneous emission models, and re-deriving asymptotic bounds for dephasing and lossy interferometer with unprecedented simplicity.
\begin{figure}[t]
 \includegraphics[width=0.5\textwidth]{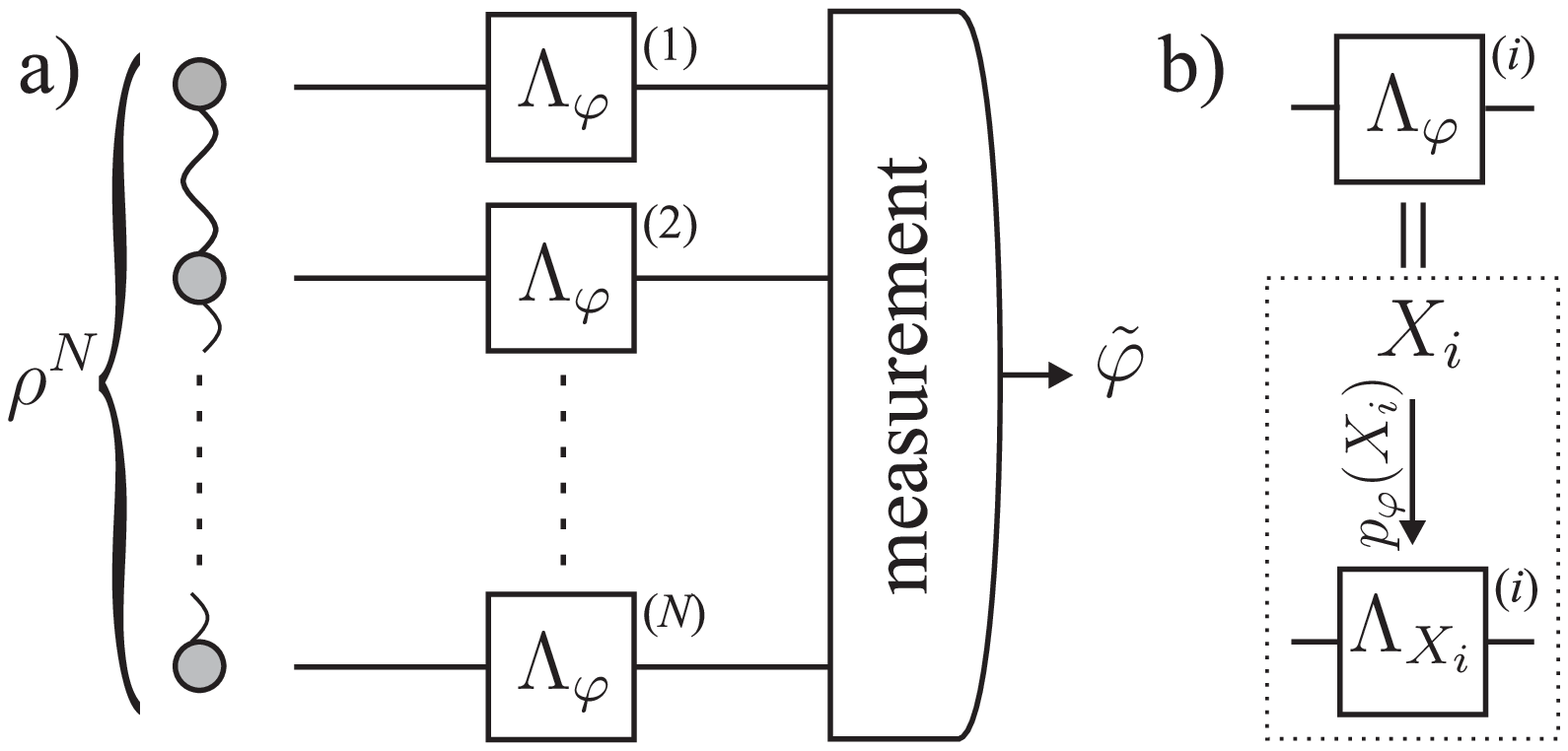}
\caption{\textsf{\textbf{Quantum metrology and the classical simulation idea.}
\textbf{a}, General scheme for quantum enhanced metrology. $N$-probe quantum state fed into $N$ parallel channels is sensing an unknown channel parameter $\varphi$. An estimator $\tilde{\varphi}$ is inferred from a measurement result on the output state.
\textbf{b}, Classical simulation of a quantum channel. The channel $\Lambda_{\varphi}$ is interpreted as a mixture of other channels $\Lambda_{X}$, where the dependence on $\varphi$ is moved into the mixing probabilities $p_{\varphi}(\! X\!)$.
}}
\label{fig:1}
\end{figure}

\section*{Bounds on precision in quantum enhanced metrology}

The typical quantum metrology scenario is as follows (see Fig.~\ref{fig:1}a). An ensemble of $N$ quantum systems
undergo in parallel the same transformation  $\Lambda_{\varphi}$, which depends on an unknown physical parameter $\varphi$. The output state is measured and the outcome is used to compute an estimate $\tilde{\varphi}$ of the parameter $\varphi$ as summarised below
\begin{equation}
\varphi\longrightarrow\Lambda_{\varphi}^{\otimes N}\!\left[\rho^{N}\right]\longrightarrow\tilde{\varphi}\label{eq:MarkovOrig}.
\end{equation}
The task is to find the optimal (possibly highly entangled) input state $\rho^{N}$ \emph{and} the most effective measurement strategy in order to minimize the estimation error $\Delta\varphi_{N}$.
Note that since the decoherence process is assumed to act independently on each of the probes, the global channel is described by the tensor product $\Lambda_\varphi^{\otimes N}$.

%
%

We pursue the estimation problem by restricting our attention to estimators which are unbiased in the neighbourhood of some fixed parameter value, for which the quantum Cram\'{e}r-Rao bound  holds
\begin{equation}\label{eq.cr}
\Delta\varphi_{N}\geq \frac{1}{\sqrt{F_{Q}\!\left[\,\Lambda_{\varphi}^{\otimes N}\!\left[\rho^{N}\right]\,\right]}},
\end{equation}
where $F_{Q}$ is the \emph{quantum Fisher information} (QFI) \cite{Braunstein1994}.
%
%
Maximization of QFI over input states $\rho^{N}$ sets the limit on the
achievable quantum enhanced precision:
\begin{equation}\label{eq:quantumFisher}
\Delta\varphi_{N}\geq\frac{1}{\sqrt{\mathcal{F}_{N}}},\
\mathcal{F}_{N}=\underset{\rho^{N}}{\max}\; F_{Q}\!\left[\,\Lambda_{\varphi}^{\otimes N}\!\left[\rho^{N}\right]\,\right].
\end{equation}
The states that maximize the QFI and yield the HS for decoherence-free unitary channels are typically highly entangled: the GHZ state in the case of atomic spectroscopy \cite{Bollinger1996}, and the N00N state in the case of optical interferometry \cite{Dowling1998}. In the presence of decoherence, the optimal input states do not have an intuitive form and the maximization of the QFI (with rising $N$) becomes hard even numerically \cite{Demkowicz2009a,Dorner2008,Huelga1997}.

\begin{figure}[t]
 \includegraphics[width=0.5 \textwidth]{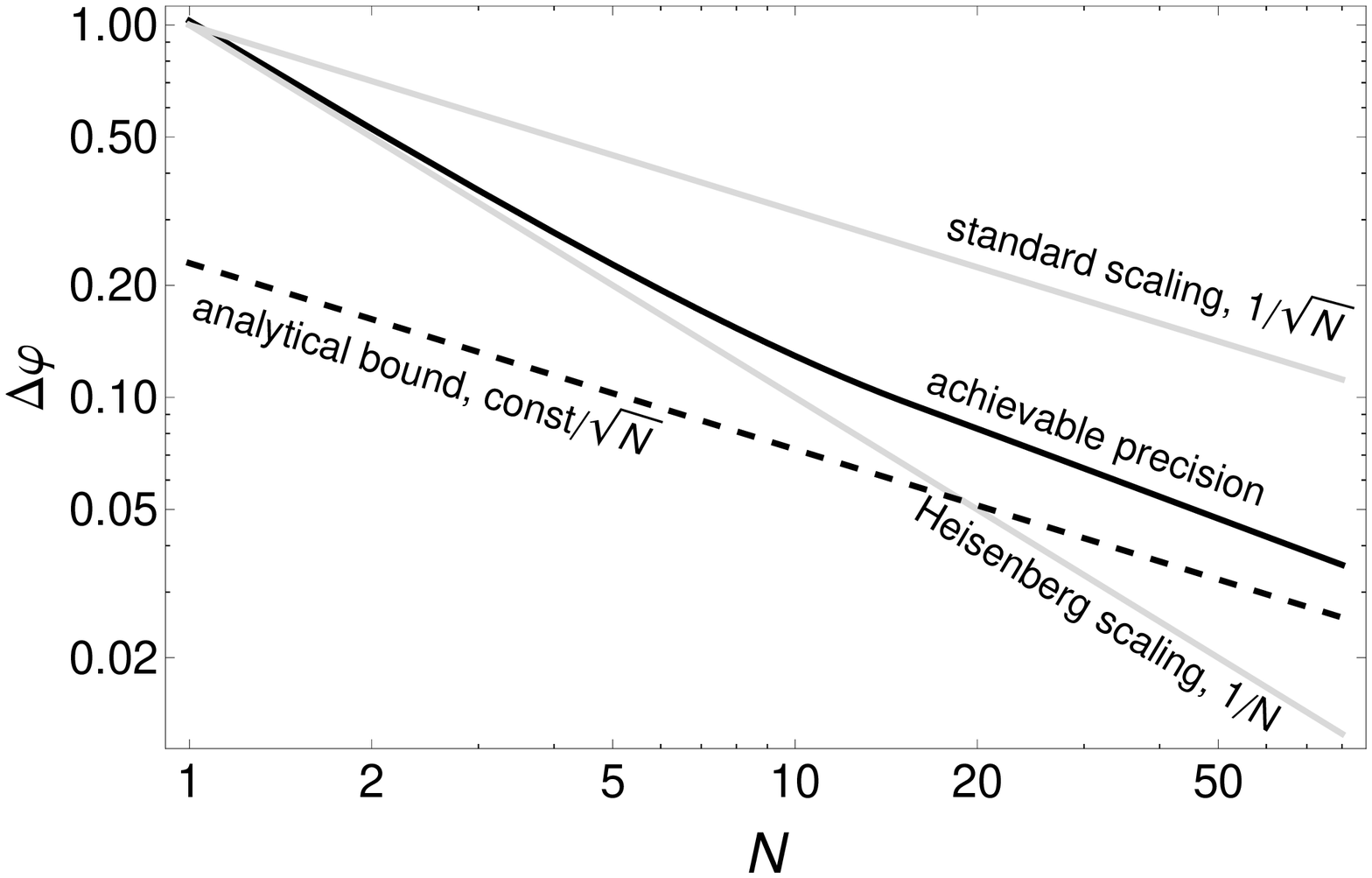}
\caption{\textsf{\textbf{Estimation precision in presence of decoherence.} Log-log plot of a generic dependence of quantum enhanced parameter estimation uncertainty in the presence of decoherence as a function of the number of probes used. While for small number of probes the curve for achievable precision follows the Heisenberg scaling,
it asymptotically flattens to approach $\t{const}/\sqrt{N}$ dependence. The ``const'' represents the quantum enhancement factor.
The exemplary data correspond to the case of phase estimation using $N$ photons in a Mach-Zehnder interferometer with $5\%$ losses in both arms.}}
\label{fig:plot}
\end{figure}
The typical behaviour of the estimation uncertainty in the presence of decoherence is depicted in the log-log scale in Fig.~\ref{fig:plot}, showing that asymptotically in $N$ the quantum gain amounts to a constant factor improvement
over the standard $1/\sqrt{N}$ scaling achievable with independent probes.
The key result of this paper is to provide a method for a general and simple calculation of this constant factor improvement.
%

\section*{Classical simulation}
To understand the idea of CS, we need to think of quantum channels in a geometrical way \cite{Bengtsson2006}. A quantum channel is a completely positive, trace preserving map acting on density matrices. The space of all such  transformations is convex:   if $\Lambda, \Lambda^{\prime}$ are two channels, then
$p \Lambda + (1-p) \Lambda^{\prime}$ can be realised by randomly applying $\Lambda$ or
$\Lambda^{\prime}$ with probabilities $p$ and $1-p$. The channels that cannot be decomposed into a convex combination of different channels (e.g. the unitary transformations) are called extremal. Note that while all interior points of the space of quantum channels are non-extremal, the boundary contains both the extremal as well as some non-extremal channels.



We say that the family $\Lambda_{\varphi}$ is \emph{classically simulated} \cite{Matsumoto2010} if each channel
is written as a classical mixture of the form
\begin{equation}
\Lambda_{\varphi}[\rho]=\!\!\int \!\! dx\; p_{\varphi}\!\left(x\right)\,\Lambda_{x}[\rho],\label{eq:contClSim}
\end{equation}
where the unknown parameter enters only through the probability distribution $p_{\varphi}$ of a random variable $X$ that indicates which channel to pick
from the set $\left\{ \Lambda_{x}\!\right\}$ (see Fig.~\ref{fig:1}b). If $X_{1},\dots, X_{N}$ are independent \emph{hidden} random variables used to generate the parallel channels we can rewrite the estimation problem as
\begin{equation*}
\varphi
\rightarrow\left\{ X_{i}\right\} _{i=1}^{N}\rightarrow\otimes_{i=1}^{N}\,\Lambda_{X_{i}}\!\left[\rho^{N}\right]\rightarrow\tilde{\varphi}.
\end{equation*}
Since conditionally on the values of  $X_{i}$ the output state does not carry any information about $\phi$, the estimation precision in the above scenario is at most equal to that of the classical problem of estimating $\phi$ given $N$ independent samples from the distribution $p_{\phi}$
\begin{equation*}
\varphi\rightarrow\left\{ X_{i}\right\} _{i=1}^{N}\rightarrow\tilde{\varphi}.
\end{equation*}
Hence, using the classical Cram{\'e}r-Rao bound (CRB) \cite{Kay1993} we obtain a lower bound on the uncertainty of the original problem \eqref{eq:MarkovOrig}
\begin{equation}
\Delta\varphi_{N}\ge\frac{1}{\sqrt{N\, F_{\t{cl}}[p_{\varphi}]}},\;\ F_{\t{cl}}=\!\!\int\!\!
dx\,\frac{\left[\partial_{\varphi}{p}_{\varphi}(x)\right]^{2}}{p_{\varphi}(x)},\label{eq:boundClSim}
\end{equation}
where $F_{\t{cl}}$ is the classical Fisher information;
see Methods for an alternative derivation. 

Provided that $p_{\varphi}(x)$ satisfies regularity conditions required in the derivation of CRB \cite{Kay1993},
the SS of precision follows immediately.
Moreover, any particular classical simulation yields an explicit SS bound on precision. If many CSs are possible,
we obtain the tightest SS bound for $\Delta\varphi_{N}$ by choosing the ``worst'' decomposition yielding minimal $F_{\t{cl}}[p_{\varphi}]$, as shown below. On the other hand, HS is possible only when the above conditions are \emph{not} satisfied.
This happens in the decoherence-free case where $\Lambda_\varphi$ are unitary channels which are extremal points of the SQC and
 the only admissible $p_{\varphi}$ in \eqref{eq:contClSim} is the Dirac delta distribution being zero on all channels except $\Lambda_{\varphi}$. This yields $F_{\t{cl}}\!=\!\infty$, makes the bound trivial and leaves the possibility of HS, which is consistent with results of decoherence-free metrology.


\begin{figure}
\includegraphics[width=0.5\textwidth]{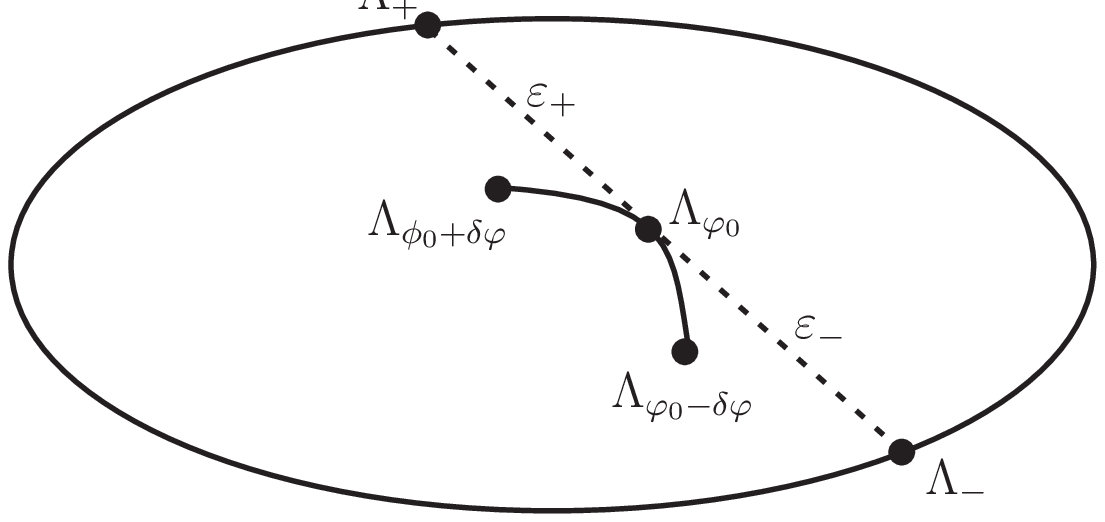}
\caption{\textsf{\textbf{Local classical simulation.} Schematic representation of a local classical simulation of a channel
that lies \emph{inside} the convex set of quantum channels.}}
\label{fig:2}
\end{figure}
The QFI at a given $\varphi_{0}$ depends only on the output state and its first derivative at $\varphi_{0}$. This implies that any family of channels $\tilde{\Lambda}_{\varphi}$ which ``locally coincides'' with the original one, i.e. 
$$
\tilde{\Lambda}_{\varphi_{0}}[\rho]\!=\!\Lambda_{\varphi_{0}}[\rho], \qquad \partial_{\varphi}\tilde{\Lambda}_{\varphi}\!\!\left.[\rho]\right|_{\varphi_{0}}\!\!=\!\partial_{\varphi}\Lambda_{\varphi}\!\!\left.[\rho]\right|_{\varphi_{0}},
$$
achieves the same maximum QFI in \eqref{eq:quantumFisher}.
It is therefore enough to consider the \emph{local classical simulations}, i.e. any mixtures reproducing the channel and its first derivative at given $\varphi_{0}$.
As proven in Supplementary Methods, Section I, the CS with the smallest $F_{\t{cl}}$  can be constructed using two channels, $\left\{ \Lambda_{+},\Lambda_{-}\right\} $, that lie on the tangent line to the {}``channel trajectory'' at the two outermost points situated on the boundary of the set of channels, see Fig~\ref{fig:2}. The local CS around $\phi_{0}$ reads explicitly
\begin{equation}
\tilde{\Lambda}_{\varphi}[\rho]=p_{\varphi}^{+}\,\Lambda_{+}[\rho]+p_{\varphi}^{-}\,\Lambda_{-}[\rho],
\end{equation}
with
$$
\Lambda_{\pm}[\rho]\!=\!\Lambda_{\varphi_{0}}[\rho]\pm\varepsilon_{\pm}\!\left.\partial_{\varphi}\Lambda_{\varphi}[\rho]\right|_{\varphi_{0}},
\quad
p_{\varphi}^{\pm}\!=\!\frac{\varepsilon_{\mp}\pm(\varphi-\varphi_{0})}{\varepsilon_{+}+\varepsilon_{-}}.
$$
Making use of Eq.~\eqref{eq:boundClSim} applied to the binary probability distribution $p_{\varphi}^{\pm}$, we obtain
\begin{equation}
\Delta\varphi_{N}\ge \sqrt{\frac{\varepsilon_{+}\varepsilon_{-}}{N}}.\label{eq:boundTanClSim}
\end{equation}
To calculate the above bound it suffices to find the ``distances'' $\varepsilon_{\pm}$ of the channel from the boundary measured along the tangent line. For extremal channels $\varepsilon_{\pm}=0$ and the bound is not useful.
For non-extremal channels,  the above construction will yield a finite $F_{\t{cl}}$ provided that $\varepsilon_{\pm}>0$, i.e.
$\Lambda_{\varphi_{0}}$ can be decomposed into a mixture of channels lying on the tangent. Channels which have this additional property will be called \emph{$\varphi$-non-extremal}. They obey the standard precision scaling, and include  all full rank channels (i.e. channels lying in the interior of the SQC) \cite{Fujiwara2008,Matsumoto2010}.
An explicit method for calculating $\varepsilon_\pm$ and hence the bound for a general quantum channel is described in the Methods section.

\section*{Channel extension}
Even though the CS method is very general, there are interesting examples
of $\varphi$-extremal decoherence models for which CS does not apply. In this case one can resort to the
more powerful but less intuitive CE method.

The action of the quantum channel $\Lambda_\varphi$
can be described via its Kraus representation \cite{Nielsen2000}:
\begin{equation*}
\Lambda_{\varphi}[\rho]  =\sum_{i}K_{i}\!\left(\varphi\right)\rho\, K_{i}^{\dagger}\!\left(\varphi\right),
\end{equation*}
with Kraus operators satisfying $\sum_i\! K_i(\varphi)^\dagger K_i(\varphi)\!=\! \openone$. Although this representation is not unique, different sets of linearly independent Kraus operators are related by unitary transformations
\begin{equation}
\tilde{K}_{i}(\varphi)=\sum_{j}\mathbf{u}_{ij}(\varphi)\, K_{j}(\varphi)
\label{eq:KrausReps},
\end{equation}
where 
$ \mathbf{u}(\varphi)$ is a unitary matrix depending on $\varphi$.

An equivalent definition of the QFI has been proposed \cite{Fujiwara2008,Escher2011}
\begin{equation}
F_{Q}\!\left[\Lambda_{\varphi}\left[\rho\right]\right]=\min_{\ket{\Psi_{\varphi}}}\,F_{Q}\!\left(\ket{\Psi_{\varphi}}\right)\!,\label{eq:minpure}
\end{equation}
where the minimization is performed over all $\varphi$ differentiable
purifications of the output state
$$
\Lambda_{\varphi}[\rho]\!=\!\t{Tr}_{E}\!\left\{ \ket{\Psi_{\varphi}}\!\bra{\Psi_{\varphi}}\right\}.
$$
For pure input state, different purifications correspond to different equivalent Kraus representations of the channel, as in Eq.~(\ref{eq:KrausReps}). For many quantum metrological models \cite{Escher2011} one can make an
educated guess of a purification and derive excellent input independent analytical bounds providing the correct asymptotic scaling of precision.
Nevertheless, the method may be cumbersome especially when the channel description involves many Kraus operators.

A simpler bound can be derived by exploiting the intuitive observation that allowing the channel to act in a trivial way on an \emph{extended space}, can only improve the precision of estimation, i.e.
$$
\max_{\rho}F_{Q}\left[\Lambda_{\varphi}\left[\rho\right]\right]\!\le\!\max_{\rho_{\t{ext}}}F_{Q}\left[\Lambda_{\varphi}\otimes\mathbb{I}\left[\rho_{\t{ext}}\right]\right].
$$
This leads to an upper bound on $\mathcal{F}_{N}$, 
which goes around the input state optimization, yielding \cite{Fujiwara2008}:
\begin{equation}
\label{eq:bound}
\mathcal{F}_{N}\leq4\min_{\tilde{K}}\left\{ N\left\Vert \alpha_{\tilde{K}}\right\Vert +N\left(N-1\right)\left\Vert \beta_{\tilde{K}}\right\Vert ^{2}\right\}
\end{equation}
 where $\left\Vert \cdot\right\Vert $ denotes the operator norm, the minimization is performed over all equivalent Kraus representations of $\Lambda_{\varphi}$, and
 \begin{equation}
\alpha_{\tilde{K}} \!=\!  \sum_{i}\dot{\tilde{K}}_{i}^{\dagger}\!(\varphi)\dot{\tilde{K}}_{i}(\varphi),\
 \beta_{\tilde{K}} \!=\!  \mathrm{i}\sum_{i}\dot{\tilde{K}}_{i}^{\dagger}\!(\varphi)\tilde{K}_{i}(\varphi),
 \end{equation}
 %
where  $\dot{\tilde{K}}_{i}(\varphi) \!=\!  \partial_{\varphi}\tilde{K}_{i}(\varphi)$.
 For any given $\varphi=\varphi_0$, equation ~\eqref{eq:bound} involves only $\tilde{K}_i$ and its first derivatives at $\varphi_0$. Moreover, the bound is insensitive to changing the Kraus representations with a $\varphi$ independent
  $\mathbf{u}$. Therefore,  it is enough to parameterize equivalent Kraus representations in
 Eq.~(\ref{eq:KrausReps}) with a hermitian matrix $\mathbf{h}$ which is generator of
 $\mathbf{u}(\varphi)=\exp[-\mathrm{i}\mathbf{h}(\varphi-\varphi_0)]$. This reduces the optimization problem \eqref{eq:bound} to a minimization over $\mathbf{h}$.

Since the SS of precision holds when $\mathcal{F}_{N}$ scales linearly with $N$, the bound \eqref{eq:bound} implies that a sufficient condition for SS is to find $\mathbf{h}$ for which $\beta_{\tilde{K}}=0$, or equivalently\cite{Fujiwara2008}:
\begin{equation}
\sum_{i,j}\mathbf{h}_{ij}K_{i}^{\dagger}K_{j}=\mathrm{i}\sum_{q}\dot{K}_{q}^{\dagger}K_{q}.\label{eq:fic}
\end{equation}
Here we go a step further and show that in this case we can obtain a quantitative SS bound
\begin{equation}\label{eq:fujibound}
\mathcal{F}_{N}\leq4N\min_{\mathbf{h}}\left\Vert \alpha_{\tilde{K}}\right\Vert,
\end{equation}
where the minimisation runs over $\mathbf{h}$ satisfying \eqref{eq:fic}, and can be formulated as  a semi-definite program, as described in the Methods section.

Moreover, it turns out that the bound resulting from finding the global minimum in Eq.~\eqref{eq:fujibound}
is at least as tight as the one derived using the CS method based on Eq.~\eqref{eq:boundClSim} (see Supplementary Methods, Section II)
proving superiority of the CE over the CS method.



\section*{Examples}
\begin{figure*}
\includegraphics[width=1 \textwidth]{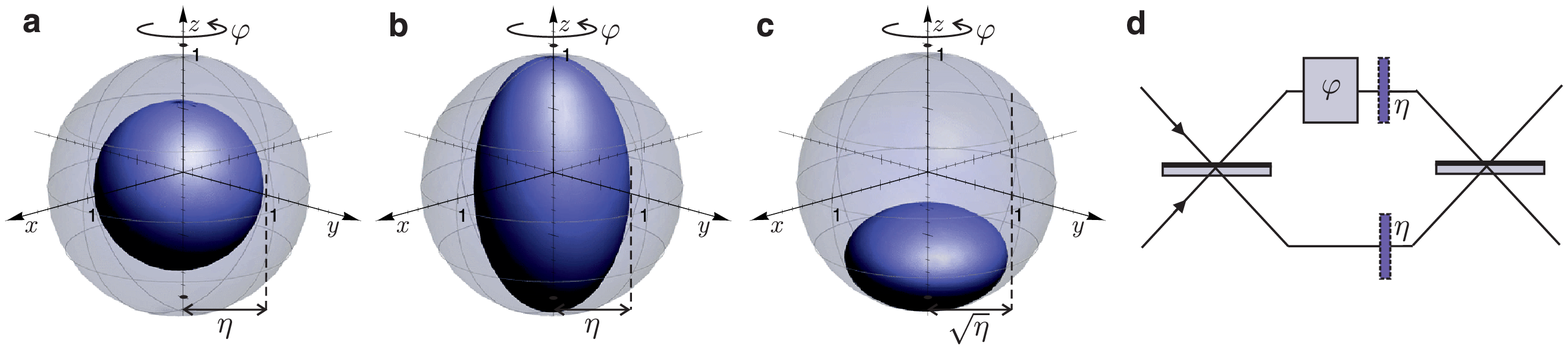}
\caption{\textsf{\textbf{Decoherence models.} Graphical representation of decoherence models discussed in the paper.
Two-level atom decoherence processes are illustrated with
a corresponding shrinking of the Bloch ball. \textbf{a}, Depolarisation \textbf{b}, Dephasing \textbf{c},
Spontaneous emission \textbf{d)} Lossy interferometer with $\eta$ being the power transmission coefficient.
}}
\label{fig:balls}
\end{figure*}

All examples of channels presented below are of the form $\Lambda_{\varphi}[\rho]\!=\!\Lambda\!\left[U_{\varphi}\rho U_{\varphi}^{\dagger}\right]$, i.e. a concatenation of a unitary rotation encoding the estimate parameter $\varphi$
and an $\varphi$-independent decoherence process. Consequently, $K_{i}(\varphi)\!=\!K_{i}U_{\varphi}$, where $K_{i}(\varphi)$,
$K_{i}$ are the Kraus operators of $\Lambda_{\varphi}$ and $\Lambda$ respectively. The most relevant models in quantum enhanced
metrology belong to this class, but the methods presented may be applied to more general models as well.

In what follows, we adopt the standard notation where $\openone$ is the $2\times2$ identity matrix and $\left\{ \sigma_{i}\right\} _{i=1}^{3}$ are the Pauli operators. We focus on two-level probe systems (qubits) sensing a phase shift modeled using a unitary $U\!(\varphi)=\exp[(\mathrm{i}\sigma_{3}\varphi)/2]$ --- rotation of the Bloch ball around the $z$ axis. In the case of atomic clocks' frequency calibration $\varphi\!=\!\delta\omega\!\cdot\! t$ with $\delta\omega$ being the detuning between the frequency of the
atomic transition and the frequency of driving field, while $t$ the time of evolution. Even though in practice the parameter to be estimated is $\delta\omega$, we will consider that to be $\varphi$, in order to have a unified notation for atomic and optical models. In the case of a two mode optical interferometer, $U\!(\varphi)$ is the operator acting on a single photon state, accounting for the accumulated relative phase shift $\varphi$ between the two arms of the interferometer.


We apply the methods to four decoherence processes encountered in quantum enhanced metrology:
two-level atom dephasing, depolarisation, spontaneous emission and the photon loss inside the interferometer.
These examples will provide us with the full picture of the applicability of the methods discussed in the paper,
as their cover all distinct cases from the point of view of the geometry of SQC.

\textbf{Depolarization.}
Two level atom depolarization describes an isotropic loss of coherence and
may be visualized by a uniform Bloch ball shrinking, see Fig.~\ref{fig:balls}a,
where $ 0\! \leq \!\eta\! <\! 1$ is the final Bloch ball radius.
Its description involves four Kraus operators
\begin{equation}
K_{0}=\sqrt{\frac{1+3\eta}{4}}\,\openone, \quad \left\{ K_{i}=\sqrt{\frac{1-\eta}{4}}\,\sigma_{i}\right\} _{i=1\dots3}
\end{equation}
which makes it an example of a channel lying in the interior of the SQC.
Using CS method we infer the SS of precision, and calculate the ``distances''
from the boundary of the SQC $\varepsilon_\pm \!=\!\sqrt{(1-\eta )(1+3\eta )}/(2\eta)$ (see Methods) resulting in the bound given in
left column of Tab.~\ref{tab:comparison}a.
Applying the CE method it is possible to further improve the bound and the result of the semi-definite optimization
is shown in the right column of Tab.~\ref{tab:comparison}a. The only non zero elements of the corresponding optimal
$\mathbf{h}$ are $\mathbf{h}_{03}=\mathbf{h}_{30}=\sqrt{(1-\eta)(1+3 \eta)}/c$ and $\mathbf{h}_{12}=-\mathbf{h}_{21}=-\mathrm{i}(1+\eta)/c$
where $c=2(1-\eta)(1+2\eta)$.
To our best knowledge, the bound has not been derived before.
%

\begin{table}[t]
\begin{tabular}{|>{\flushleft}m{0.5cm}>{\centering}m{2.5cm}|>{\centering}m{2.5cm}|>{\centering}m{2.5cm}|}
\hline
\multicolumn{2}{|>{\centering}m{3cm}|}{\textsf{\textbf{channel considered}}}  & \textsf{\textbf{classical simulation}} \eqref{eq:boundTanClSim}  & \textsf{\textbf{channel extension}} \eqref{eq:fujibound}  \tabularnewline
\hline
\hline
\noalign{\vskip\doublerulesep}
~

\textsf{\textbf{ a}}

~
&
~

\textsf{depolarisation}

~
&
~
$\sqrt{\frac{(1-\eta)(1+3\eta)}{4 \eta^2}}$
~
&
\multicolumn{1}{>{\centering}m{2.5cm}|}{$\sqrt{\frac{1+\eta-2\eta^2}{2 \eta^2}}$}\tabularnewline
\hline
~

\textsf{\textbf{ b}}

~
&
~

\textsf{dephasing}

~ &  \multicolumn{2}{>{\centering}m{5cm}|}{$\frac{\sqrt{1-\eta ^2}}{\eta }$}\tabularnewline
\hline
~

\textsf{\textbf{ c}}

~
&

\textsf{spontaneous emission}  & ~

N/A

~ & \multicolumn{1}{>{\centering}m{2.5cm}|}{$\frac{1}{2}\sqrt{\frac{1-\eta}{\eta}}$ }\tabularnewline
\hline
~

\textsf{\textbf{ d}}

~
&
\textsf{lossy interferometer} & ~

N/A

~ & \multicolumn{1}{>{\centering}m{2.5cm}|}{$\sqrt{\frac{1-\eta}{\eta}}$ }\tabularnewline
\hline
\end{tabular} \caption{\textsf{Precision bounds of the most relevant models in quantum enhanced metrology derived with two methods discussed in the paper. All the bounds are of the form $\Delta\varphi_{N}\geq\frac{\t{const}}{\sqrt{N}}$, where constant factors are given in the table. Classical simulation method does not provide bounds for spontaneous emission and lossy interferometer, since these channels are $\varphi$-extremal. For the dephasing model it surprisingly yields an equally tight bound as the more powerful channel extension method.}}
\label{tab:comparison}
\end{table}
\textbf{Dephasing.}
Dephasing is a decoherence model of two-level atoms subject to fluctuating external magnetic/laser fields.
In graphical representation, it corresponds to shrinking of the Bloch ball in $x,y$ directions with $z$ direction intact, see
Fig.~\ref{fig:balls}b. The canonical Kraus operators read \cite{Nielsen2000}
\begin{equation}
K_{0}=\sqrt{\frac{1+\eta}{2}}\,\openone,\quad K_{1}=\sqrt{\frac{1-\eta}{2}}\,\sigma_{3},
\end{equation}
where $0\!\leq\! \eta \!<\! 1$ is the dephasing parameter.
As it involves only two Kraus operators, it is not a full-rank channel and lies on the boundary of the SQC.
It is, however, a non-extremal and more importantly non-$\varphi$-extremal channel allowing for the CS
construction with $\varepsilon_\pm\!=\!\sqrt{1-\eta ^2}/\eta$ (see Methods), yielding the bound given in Tab.~\ref{tab:comparison}b.

Most importantly, the above bound, is exactly the same as the one derived by Escher \emph{et al.}\cite{Escher2011} or by
the CE method, where the minimum in Eq.~\eqref{eq:fujibound} corresponds to
$\mathbf{h}\!=\!\sigma_{1}/(2\sqrt{1-\eta^2})$.
This proves that despite its simplicity the CS method may sometimes lead to bounds that are equally tight as the ones derived with much more advanced methods even for channels lying on the boundary of the SQC.

\textbf{Spontaneous emission.}
The well known two-level atom spontaneous emission model is described by the Kraus operators
\begin{equation}
K_{0}=\left(\begin{array}{cc}
1 & 0\\
0 & \sqrt{\eta}\end{array}\right),\: K_{1}=\left(\begin{array}{cc}
0 & \sqrt{1-\eta}\\
0 & 0\end{array}\right)
\end{equation}
with $0\!\leq \!\eta \!<\! 1$. Interestingly, for all $\eta$ this channel is extremal \cite{Ruskai2001}, which means that the CS is
not applicable. Nevertheless, the CE method of Eq.~\eqref{eq:fujibound}, can be easily employed.

Substituting $K_{i}(\varphi)$ into Eq.\eqref{eq:fic} we find that the generator $\mathbf{h}$ is fixed to
$\mathbf{h} = \frac{1}{2(1-\eta)}(\sigma_z - \eta \openone)$,
Consequently, there is no need for minimization over $\mathbf{h}$ and from Eq.~\eqref{eq:fujibound} we automatically obtain an SS bound listed in
Tab.~\ref{tab:comparison}c which to our knowledge has not been reported in the literature before.


\textbf{Lossy interferometer.}
In order to model \emph{loss in an optical interferometer}, a third orthogonal state at the output, --- vacuum --- resulting from a loss of a photon, needs to be introduced. The decoherence channel on a single probe (single photon) is a map from a two to a three dimensional system:
\begin{equation}
\begin{array}{c}
K_{0}\!=\!\left(\begin{array}{cc}
0 & 0\\
0 & 0\\
0 & \sqrt{1-\eta}\end{array}\right)\!,\; K_{1}\!=\!\left(\begin{array}{cc}
0 & 0\\
0 & 0\\
\sqrt{1-\eta} & 0\end{array}\right)\!,\\
K_{2}\!=\!\left(\begin{array}{cc}
\sqrt{\eta} & 0\\
0 & \sqrt{\eta}\\
0 & 0\end{array}\right)\!,\end{array}
\end{equation}
where $\eta$ is the power transmission coefficient for the light traveling through each of the two arms.
Although the corresponding channel $\Lambda_{\varphi}$ is non-extremal, it is unfortunately $\varphi$-extremal and CS cannot be used.
Still, CE method can be easily applied. 
The optimal bound (see Tab.~\ref{tab:comparison}d) corresponds to $\mathbf{h}$ with non-zero elements
$\mathbf{h}_{00}\!=-\mathbf{h}_{11}=\!\frac{1}{2(1-\eta)}$.
This bound is asymptotically equally tight to the best bounds known in the literature \cite{Escher2011,Kolodynski2010,Knysh2011} proving again that
CE method despite its simplicity is able to provide powerful results in a straightforward manner.


In optical interferometric applications it is common to use states of light with an unbounded number of photons such as coherent or squeezed states. At a first glance it is not obvious that the model considered in the paper covers these situations. Formally speaking the bound we have derived applies to input states with a total number of photons fixed to $N$. Notice, however,
that in every optical experiment what is measured in the end are photon numbers. If all phase reference beams are taken into account
we can regard quantum states of light as incoherent mixtures of states occupying different total photon number sectors \cite{Molmer1996, Jarzyna2012}.
From the point of view of metrology the QFI is then bounded from above by the weighted sum of QFIs for each of these sectors \cite{Demkowicz2009a}.
Within each sector we can easily apply our bound and since the bound is linear in $N$ the effective bound will simply correspond to replacing
$N$ with $\bar{N}$ --- mean number of photons in all relevant beams used in the
experiment --- and as such can be automatically applied to experiments involving coherent and squeezed light.


In order to demonstrate the practical relevance of the bound it is instructive to compare its predictions with the actual quantum enhancement
observed recently in the GEO600 gravitational wave detector \cite{LIGO2011}.
A detailed theoretical analysis of the setup from the perspective of the derived bounds is underway.
Still, by reducing the essential features of the setup to a simple
Mach-Zehnder interferometer, we can give a preliminary estimate on how far is the actual experiment from the optimal performance.
For the reported overall optical transmission $\eta=0.62$ the theoretically predicted maximal quantum enhancement
amounts to a  $\sqrt{1-\eta}=0.62$ factor reduction in estimation uncertainty compared with the classical $1/\sqrt{\eta N}$ limit.
The reported experimentally observed reduction was a factor of $0.67$ which is an indication that the experiment operates close
to the fundamental quantum limit and any significant improvement is possible only if the optical loss is further reduced.

\section*{Summary}
Assessing the impact of decoherence on the maximum possible quantum enhancement is a
crucial element in developing quantum techniques for metrological applications.
The tools developed in the paper allow for a direct calculation of bounds on the precision
enhancement for arbitrary parameter estimation model where the decoherence process acts independently on each of the probes and may be represented with a finite
number of Kraus operators. While the CE method is more powerful
and for the most relevant metrological models yields in the asymptotic limit of large number of probes the tightest bounds known in the literature, the CS method may fail to provide equally tight bounds but provides an intuitive geometric insight into the absence of asymptotic HS in the presence of decoherence.
From the derived bounds it is clear that if the HS was to be preserved for large number of probes $N$
the level of decoherence would have to decrease with increasing $N$ roughly as $(1-\eta) \approx 1/N$.
This gives an estimate on the regime in which the quantum enhancement is quadratic as compared to the regime of constant factor improvement. This is clearly seen in Fig.~\ref{fig:2} where this transition appears around $N \approx 1/(1-\eta) = 20$ for $\eta=0.95$. Since in most metrology applications $N$ is larger by several order of magnitude, we expect that the SS scaling provides a reliable bound for the optimal estimation precision.

An important question that was not addressed in the paper is the saturability of the bounds. Clearly the CS method
does not provide tight bounds in general as we have observed e.g. for depolarization channel.
As to the CE method we are not able to prove that the bounds we have derived are saturable in the asymptotic limit.
We should stress however, that these are lower bounds on estimation uncertainties and an upper bound can always be
found by choosing a particular estimation method. Therefore if a particular strategy performs close to the derived bound, we can certify that it is near the fundamental quantum limit, as illustrated by the results of GEO600 experiment.

Note that even though
the minimization over purification method\cite{Escher2011} yields in principle a tight bound,
in practice there is no effective algorithm to find a global minimum
and the only way to convince oneself that the one has hit the global minimum is again to show that the theoretical lower bound on estimation uncertainty
 coincides with the performance of a particular estimation strategy.

The methods discussed were focused on deriving useful bounds in the asymptotic regime of large number of probes.
Still, the bounds are valid (though weaker) for any value $N$. We leave it for a future work to improve the bounds
for finite $N$ which seems to be possible by using the CE method and relaxing the $\beta_{\tilde{K}}=0$ constraint.

\section*{Methods}
{\small
\textbf{Proof of the classical simulation bound.}
In order to simplify the reasoning, let us focus on the classical simulations that are constructed using discrete
sets of quantum channels, $\left\{ \Lambda_{i}\right\}$. Then, the considered channel's action of Eq.~(\ref{eq:contClSim}) can be rewritten as a $\varphi$-independent map acting on a larger input space \cite{Matsumoto2010}
\begin{equation}
\Lambda_{\varphi}[\rho]=\sum_{i}\, p_{\varphi,i}\,\Lambda_{i}[\rho]=\Phi\left[\rho\otimes\sigma_{\varphi}\right],\label{eq:discClSim}
\end{equation}
where $\sigma_{\varphi}\!=\!\sum_{i}\, p_{\varphi,i}\left|e_{i}\right>\!\!\left<e_{i}\right|$ represents a diagonal state in a basis, in which $\Phi$ is defined via $\Phi[\varrho]\!=\!\sum_{i}\Lambda_{i}\!\otimes\! E_{i}\,[\varrho]$
with $E_{i}[\sigma]=\left<e_{i}\right|\sigma\left|e_{i}\right>$.
In order to prove \eqref{eq:boundClSim}, we write the QFI for the
$N$ parallel use of the channel and bound it from above, i.e.
\begin{align}
F_{Q}\!\left[\Lambda_{\varphi}^{\otimes N}[\rho]\right] & =F_{Q}\!\left[\Phi^{\otimes N}\left[\rho\otimes\sigma_{\varphi}^{\otimes N}\right]\right]\nonumber \\
 & \le F_{Q}\!\left[\rho\otimes\sigma_{\varphi}^{\otimes N}\right]=F_{Q}\!\left[\sigma_{\varphi}^{\otimes N}\right]\nonumber \\
 & =N\, F_{Q}\!\left[\sigma_{\varphi}\right]=N\, F_{\t{cl}}\!\left[p_{\varphi}\right]\!,
\end{align}
exploiting the monotonicity of the QFI under any parameter independent
quantum map\cite{Fujiwara2001}, here $\Phi^{\otimes N}$.

\textbf{Calculation of the classical simulation bound.}
The geometry of the space of channels and more specifically the $\varphi$-extremality are best viewed by using the Choi-Jamio\l{}kowski isomorphism \cite{Jamiolkowski1972,Choi1975}. Given a quantum channel $\Lambda\!:\mathcal{L}(\mathcal{H}_{\t{in}})\!\mapsto\!\mathcal{L}(\mathcal{H}_{\t{out}})$ acting from the space of density matrices on $\mathcal{H}_{\t{in}}$ to density matrices on $\mathcal{H}_{\t{out}}$, one defines $P_{\Lambda}\!=\!\Lambda\otimes\mathbb{I}\,[\left|\Phi\right>\!\left<\Phi\right|]$, where $\left|\Phi\right>\!=\!\sum_{i=1}^{\dim\mathcal{H}_{\t{in}}}\ket{i}\!\otimes\!\ket{i}$ is a maximally entangled state in $\mathcal{H}_{\t{in}}\otimes\mathcal{H}_{\t{in}}$. $\Lambda$ is a physical channel (i.e. trace preserving, completely positive map) iff $P_{\Lambda}$ is a positive semi-definite operator, satisfying $\t{Tr}_{\mathcal{H}_{\t{out}}}\!\{P_{\Lambda}\}\!=\!\openone$. If $\{K_{i}\}_{i}$ are the Kraus operators of the $\Lambda$ channel, we can write explicitly $P_{\Lambda}=\sum_{i}\ket{K_{i}}\bra{K_{i}}$, where $\ket{K_{i}}=K_{i}\otimes\openone\ket{\Phi}$.

We can now say that, the channel $\Lambda_{\varphi}$ is \emph{not} $\varphi$-extremal, if it is possible to find a non-zero $\varepsilon$, for which $P_{\Lambda_{\varphi}}\pm\varepsilon\,\partial_{\varphi}\! P_{\Lambda_{\varphi}}\ge0$. See Supplementary Methods, Section III for an alternative formulation of the $\varphi$-extremality condition and its relation to the well known non-extremality condition due to Choi \cite{Choi1975}. In practice, if we want to make most out of the bound in Eq.~\eqref{eq:boundTanClSim}, we need to find the maximum values of $\varepsilon_{\pm}$, for which $P_{\Lambda_{\varphi}}\pm\varepsilon_{\pm}\,\partial_{\varphi}P_{\Lambda_{\varphi}}$ are still positive semi-definite operators.
This is a simple eigenvalue problem and therefore the bound can be obtained immediately.

Taking the dephasing model as an example, the Choi-Jamio{\l{}}kowski isomorphism of the corresponding $\Lambda_{\varphi}$ channel $P_{\Lambda_{\varphi}}\!=\!\sum_{i=1}^{2}\ket{K_{i}U_{\varphi}}\bra{K_{i}U_{\varphi}}$ has a simple form:
\begin{equation}
P_{\Lambda_{\varphi}}=\left(\begin{array}{cccc}
1 & 0 & 0 & \eta\,\mathrm{e}^{\mathrm{i}\varphi}\\
0 & 0 & 0 & 0\\
0 & 0 & 0 & 0\\
\eta\,\mathrm{e}^{-\mathrm{i}\varphi} & 0 & 0 & 1\end{array}\right).
\end{equation}
It is easy to check that $P_{\Lambda_{\varphi}}\!\!+\!\varepsilon\,\partial_\varphi P_{\Lambda_{\varphi}}\!\geq\!0$ provided $|\varepsilon|\!\leq\!\sqrt{1-\eta^2}/\eta$, hence using Eq.~\eqref{eq:boundTanClSim} we arrive at the bound
given in Table~\ref{tab:comparison}.

\textbf{Channel extension method as a semi-definite program.}
Here we show that the minimization problem in Eq.~\eqref{eq:fujibound} can be formulated as a simple semi-definite program.
Let the channel $\Lambda_\varphi$ be a map from a $d_1$ to a $d_2$ dimensional Hilbert spaces with Kraus representation involving $k$ linear
independent Kraus operators ($d_2 \times d_1$ matrices).
Consider the following block matrix:
\begin{equation}
A = \mat{ccccc}{
\sqrt{t} \openone_{d_1} & \dot{\tilde{K}}_0^\dagger & \dot{\tilde{K}}_1^\dagger & \dots & \dot{\tilde{K}}_{k-1}^\dagger \\
\dot{\tilde{K}}_0 & \sqrt{t} \openone_{d_2}    & 0 & \dots & 0\\
\dot{\tilde{K}}_1& 0 & \sqrt{t} \openone_{d_2}    & \dots & 0\\
\vdots & \vdots&  \vdots& \ddots& \vdots  \\
\dot{\tilde{K}}_{k-1} & 0  & 0 & \dots & \sqrt{t} \openone_{d_2}
},
\end{equation}
where $\openone_d$ is a $d \times d$ identity matrix. Positive semi-definiteness of the matrix $A$ is equivalent to
the condition
$$
\alpha_{\tilde{K}}=\sum_i \dot{\tilde{K}}^\dagger_i \dot{\tilde{K}}_i \leq t \openone_{d_1}.
$$
Minimizing the operator norm $\Vert \alpha_{\tilde{K}} \Vert$ is thus equivalent to minimizing $t$ subject to
$A \geq 0$. Taking into account Eq.~\eqref{eq:fic} the problem takes the form:
\begin{equation}
\min_{\mathbf{h}} t,\  \textrm{subject to: } A\geq 0, \sum_{ij}\mathbf{h}_{ij}K_i^\dagger K_j = \mathrm{i}\sum_q \dot{K}^\dagger_q K_q.
\end{equation}
Since $\dot{\tilde{K}}_i = \dot{K}_i - \sum_j \mathrm{i} \mathbf{h}_{ij} K_j$ the matrix $A$ is linear in $\mathbf{h}$
and the problem is thus a semi-definite program with the resulting minimal $t$ being the minimal operator norm $\Vert \alpha_{\tilde{K}} \Vert$.
For the purpose of this paper we have implemented the program using the CVX package for Matlab \cite{CVX}.
}
\acknowledgements
{\small We thank Konrad Banaszek for many inspiring discussions and constant support. R.D.-D. and J.K. were supported by the European Commission under the IP project Q-ESSENCE, ERA-NET CHIST-ERA project QUASAR and the Foundation for Polish
Science under the TEAM programme. M.G. was supported by the EPSRC Fellowship EP/E052290/1. }

\appendix
\section{Optimal tangent simulation}

We prove that the optimality of CS is reached by choosing two channels, which
are as far as possible from $\Lambda_{\varphi_{0}}$ on the tangent,
but still lie in the set of physical maps. Consider a CS being a mixture
composed of channels that lie on the tangent line to the channel trajectory.
Pick a channel $\Lambda_{0}$ situated a position $x_{0}$, $\Lambda_{0}=\Lambda_{\varphi_{0}}+x_{0}\left.\partial_{\varphi}\Lambda_{\varphi}\right|_{\varphi_{0}}$,
which contributes to the mixture with $p_{\varphi}^{0}$. Let us now
remove this channel from the mixture and replace it by two channels
$\Lambda_{\pm}$ at positions $x_{\pm}$, $x_{-}<x_{0}<x_{+}$, with
mixing probabilities $p_{\varphi}^{\pm}$. For the total probability
to sum up to one and the new mixture to be locally equivalent to the
previous one, the following quantities as well as their first derivatives
over $\varphi$ should equalize at $\varphi_{0}$: \begin{equation}
p_{\varphi}^{+}+p_{\varphi}^{-}=p_{\varphi}^{0},\quad p_{\varphi}^{+}\, x_{+}+p_{\varphi}^{-}\, x_{-}=p_{\varphi}^{0}\, x_{0}.\end{equation}
 Solving the above set of equations we get \begin{equation}
p_{\varphi_{0}}^{\pm}=p_{\varphi_{0}}^{0}\frac{x_{\pm}-x_{0}}{x_{\pm}-x_{\mp}},\quad\partial_{\varphi}p_{\varphi}^{\pm}|_{\varphi_{0}}=\partial_{\varphi}p_{\varphi}^{0}|\varphi_{0}\frac{x_{\pm}-x_{0}}{x_{\pm}-x_{\mp}}.\label{eq:condeq}\end{equation}
We now want to show that the described operation can only decrease the
Fisher information $F_{\t{cl}}$. For a moment consider a situation
in which channels $\Lambda_{\pm}$ were not used in the original mixture
and only appeared at the expense of the removed $\Lambda_{0}$ channel.
The change of $F_{\t{cl}}$ is given by \begin{equation}
\delta F_{\t{cl}}=\left.\frac{[\partial_{\varphi}p_{\varphi}^{+}]^{2}}{p_{\varphi}^{+}}+\frac{[\partial_{\varphi}p_{\varphi}^{-}]^{2}}{p_{\varphi}^{-}}-\frac{[\partial_{\varphi}p_{\varphi}^{0}]^{2}}{p_{\varphi}^{0}}\right|_{\varphi=\varphi_{0}}\end{equation}
and by Eq.~\eqref{eq:condeq} it must be zero.

If, on the other hand, channels $\Lambda_{\pm}$ were already in use
in the original mixture, we can artificially separate their initial
part from the one originated at the expense of taking away probability
from the point $x_{0}$. This formally corresponds to introducing
additional distinguishing symbol that would split the instances of
the classical random variable $x_{\pm}$ into two groups, depending
on their origin. By the same reasoning as before, the $F_{\t{cl}}$
must remain constant during probabilities redistribution. Finally,
removing the distinguishing symbol we can only reduce the $F_{\t{cl}}$,
what proves that indeed replacing $\Lambda_{0}$ with $\Lambda_{\pm}$
in a local classical simulation can only decrease the $F_{\t{cl}}$.

This reasoning may now be applied to an arbitrary probability distribution
of channels lying on the tangent line. We can only decrease the $F_{\t{cl}}$
by deleting one of the channels and redistributing its probability
to the ones lying at the line ends. Repeating this procedure for all
channels on the tangent line we conclude that the optimal (yielding
minimum $F_{\t{cl}}$) simulation is the one composed of two channels
situated at the intersection of the tangent line with the boundary
of the set of all quantum channels.

The missing point in the above reasoning is to prove that it is not
possible to decrease $F_{\t{cl}}$ further by including channels that
lie \emph{outside} the tangent line. We have proved this fact for
the class of channels $\Lambda_{\varphi}[\rho]=\Lambda(U_{\varphi}\rho U_{\varphi}^{\dagger})$
--- concatenation of a unitary parameter encoding and a decoherence
map. The sketch of the proof is the following. Notice that the values
of parameters $x_{i}$ appearing in Eq.~\eqref{eq:condeq} will now
be obtained by performing a projection of the channel onto the tangent
line. The only way, in which the use of this larger class of channels
could lead to lower $F_{\t{cl}}$, would be the situation when after
projecting a channel onto the tangent line, we find ourselves outside
the physical region of all quantum maps. The effective $x_{i}$ lies
then further apart than when considering only channels on the tangent
line. This seems like a possible way to increase the product $\varepsilon_{+}\varepsilon_{-}$.
However, taking one channel, e.g. $\Lambda_{+}$, to the {}``top''
of the tangent line requires taking the complementary channel, i.e. $\Lambda_{-}$, to the
{}``bottom''. Since the decoherence structure is $\varphi$-invariant
and the set of channels is convex, the gain in the $\varepsilon_{+}$
will be compensated by at least an equal loss in the distance $\varepsilon_{-}$,
which will \emph{not} make the product $\varepsilon_{+}\varepsilon-$
larger, and therefore will \emph{not} decrease the $F_{\t{cl}}$.

\section{\label{sec:CSimpliesFIC} Channel extension vs. classical simulation method}

If a channel can be locally simulated at $\varphi_0$, than up to the first order in $\varphi-\varphi_0$ we have
\begin{align*}
\Lambda_{\varphi}[\rho] & =p_{\varphi}^{+}\,\Lambda_{+}[\rho]+p_{\varphi}^{-}\,\Lambda_{-}[\rho]\\
 & =\sum_{i}p_{\varphi}^{+}K_{i}^{+}\rho K_{i}^{+\dagger}+\sum_{j}p_{\varphi}^{-}K_{j}^{-}\rho K_{j}^{-\dagger}\\
 & =\sum_{k}\tilde{K}_{q}\rho \tilde{K}_{q}^{\dagger},\end{align*}
where $\left\{ \tilde{K}_{q}\right\} =\left\{ \sqrt{p_{\varphi}^{+}}K_{i}^{+}\right\}\cup\left\{ \sqrt{p_{\varphi}^{-}}K_{j}^{-}\right\}$.
The constructed Kraus operators satisfy Eq.(12), since
\begin{align}
&\beta_{\tilde{K}} = \mathrm{i}\sum_{q}\dot{\tilde{K}}_{q}^{\dagger}\tilde{K}_{q}  =
\mathrm{i} \bigg(\sum_{i}\dot{p}_{\varphi}^{+}K_{i}^{+\dagger}K_{i}^{+}+ \\ &+ \sum_{j}\dot{p}_{\varphi}^{-}K_{j}^{-\dagger}K_{j}^{-}\bigg)
  =\mathrm{i}\left(\dot{p}_{\varphi}^{+}+\dot{p}_{\varphi}^{-}\right)\openone=0,\end{align}
where we have used  $\forall_{\varphi}\!:\; p_{\varphi}^{+}+p_{\varphi}^{-}=1$.

Moreover, notice that
\begin{align}
\Vert \alpha_{\tilde{K}} \Vert= \left \Vert \sum_q \dot{\tilde{K}}_q  \dot{\tilde{K}}_q \right \Vert \leq
\left(\partial_\varphi{\sqrt{p_\varphi^+}}\right)^2\left\Vert \sum_i K_i^{+\dagger} K_i^+ \right \Vert +\\
\left(\partial_\varphi{\sqrt{p_\varphi^-}}\right)^2\left\Vert \sum_i K_i^{-\dagger}
 K_i^{-} \right \Vert = \frac{[\partial p_\varphi^+]^2}{4 p_\varphi^+}+\frac{[\partial p_\varphi^-]^2}{4 p_\varphi^-}.
 \end{align}
Substituting the above inequality to Eq.~(13) 
 we see that
$\mathcal{F}_N \leq N F_\t{cl}$, where $F_{\t{cl}}$ is just the Fisher information for the local CS.
This shows that the bound derived from CS will never be tighter than the one derived from
the CE method optimized over all Kraus decompositions.

\section{\label{sec:phiExtremCond} $\varphi$ non-extremality }

We prove below that the condition on $\varphi$ non-extremality, which
requires the existence of a non-zero epsilon such that \begin{equation}
P_{\Lambda_{\varphi}}\pm\varepsilon\,\partial_{\varphi}\! P_{\Lambda_{\varphi}}\ge0,\label{eq:phiExtremCond1}\end{equation}
is equivalent to the statement that there exist a non zero Hermitian
matrix $\mu_{ij}$ such that \begin{equation}
\partial_{\varphi}P_{\Lambda_{\varphi}}=\sum_{ij}\mu_{ij}\left|K_{i}\right>\!\left<K_{j}\right|.\label{eq:phiExtremCond2}\end{equation}

Assume that \eqref{eq:phiExtremCond1} holds and recall that $P_{\Lambda_{\varphi}}=\sum_{i}\ket{K_{i}}\!\bra{K_{i}}$.
Since the operator on the l.h.s. is positive \begin{equation}
\bra{\psi}\!\left[\sum_{i}\ket{K_{i}}\!\bra{K_{i}}\pm\varepsilon\,\sum_{i}\ket{\dot{K_{i}}}\!\bra{K_{i}}+\ket{K_{i}}\!\bra{\dot{K_{i}}}\right]\!\ket{\psi}\geq0\label{eq:poscond}\end{equation}
for any vector $\ket{\psi}$. In order to prove \eqref{eq:phiExtremCond2},
it is enough to show that $\ket{\dot{K}_{j}}$ can be written as linear
combinations of $K_{i}$. If this was not the case for one of vectors,
e.g. $\ket{\dot{K}_{\bar{i}}}$, we would additionally need a vector
$\ket{L_{\bar{i}}}$ that is orthogonal to the the space spanned by
$\{\ket{K_{i}}\}_{i}$, then taking $\ket{\psi}=\sqrt{p}\ket{K_{\bar{i}}}+\exp(i\xi)\sqrt{1-p}\ket{L_{\bar{i}}}$
positivity condition \eqref{eq:poscond} leads to \begin{equation}
p|\braket{K_{\bar{i}}}{K_{\bar{i}}}|^{2}\pm\sqrt{p(1-p)}\varepsilon\left(e^{\mathrm{i}\xi}\braket{L_{\bar{i}}}{\dot{K}_{\bar{i}}}+c.c\right)\geq0\end{equation}
For any nonzero $\varepsilon$ and nonzero $\braket{L_{\hat{i}}}{\dot{K}_{\hat{i}}}$
we can always find some $\xi$ and $p$ small enough, so that the
l.h.s. is negative. This leads to a contradiction, hence \eqref{eq:phiExtremCond2}
must hold.

For the opposite direction, assume \eqref{eq:phiExtremCond2} holds
and substitute the formula for $\partial P_{\Lambda_{\varphi}}$ into
the l.h.s. of \eqref{eq:phiExtremCond2}. We need to show that there
exist $\varepsilon$ such that \begin{equation}
A^{\pm}=\sum_{i}\ket{K_{i}}\!\bra{K_{i}}\pm\varepsilon\sum_{ij}\mu_{ij}\ket{K_{i}}\!\bra{K_{j}}\geq0.\end{equation}
Defining matrices $\nu^{\pm}=\mathbf{1}\pm\varepsilon\mu$, note that,
for $\varepsilon$ small enough, they are positive semi-definite.
Hence, we can take their square root and construct $\ket{\tilde{K}_{i}^{\pm}}=\sum_{j}\left[\sqrt{\nu^{\pm}}\right]_{ij}\ket{K_{j}}$.
Then, $A^{\pm}=\sum_{i}\ket{\tilde{K}_{i}}\!\bra{\tilde{K}_{i}}$,
which is clearly positive semi-definite.$\blacksquare$

As a last remark, observe that tracing out \eqref{eq:phiExtremCond2}
over $\mathcal{H}_{\t{out}}$ space yields \begin{equation}
0=\sum_{ij}\mu_{ij}K_{i}^{\dagger}K_{j},\end{equation}
which is the Choi condition for non-extremality$^\textrm{45}$.
This reflects the trivial fact that if a channel is $\varphi$ non-extremal,
then it must \emph{not} be extremal.


\begin{thebibliography}{10}
\expandafter\ifx\csname url\endcsname\relax
  \def\url#1{\texttt{#1}}\fi
\expandafter\ifx\csname urlprefix\endcsname\relax\def\urlprefix{URL }\fi
\providecommand{\bibinfo}[2]{#2}
\providecommand{\eprint}[2][]{\url{#2}}

\bibitem{Giovannetti2011}
\bibinfo{author}{Giovannetti, V.}, \bibinfo{author}{Lloyd, S.} \&
  \bibinfo{author}{Maccone, L.}
\newblock \bibinfo{title}{Advances in quantum metrology}.
\newblock \emph{\bibinfo{journal}{Nature Photon.}}
  \textbf{\bibinfo{volume}{5}}, \bibinfo{pages}{222--229}
  (\bibinfo{year}{2011}).

\bibitem{Maccone2011}
\bibinfo{author}{Maccone, L.} \& \bibinfo{author}{Giovannetti, V.}
\newblock \bibinfo{title}{Quantum metrology: Beauty and the noisy beast}.
\newblock \emph{\bibinfo{journal}{Nature Phys.}} \textbf{\bibinfo{volume}{7}},
  \bibinfo{pages}{376--377} (\bibinfo{year}{2011}).

\bibitem{Paris2009}
\bibinfo{author}{Paris, M. G.~A.}
\newblock \bibinfo{title}{Quantum estimation for quantum technology}.
\newblock \emph{\bibinfo{journal}{Int. J. Quant. Inf.}}
  \textbf{\bibinfo{volume}{7}}, \bibinfo{pages}{125--137}
  (\bibinfo{year}{2009}).

\bibitem{Banaszek2009}
\bibinfo{author}{Banaszek, K.}, \bibinfo{author}{Demkowicz-Dobrzanski, R.} \&
  \bibinfo{author}{Walmsley, I.~A.}
\newblock \bibinfo{title}{Quantum states made to measure}.
\newblock \emph{\bibinfo{journal}{Nature Photon.}}
  \textbf{\bibinfo{volume}{3}}, \bibinfo{pages}{673--676}
  (\bibinfo{year}{2009}).

\bibitem{Dirac1958}
\bibinfo{author}{Dirac, P. A.~M.}
\newblock \emph{\bibinfo{title}{The Principles of Quantum Mechanics}}
  (\bibinfo{publisher}{Oxford University Press}, \bibinfo{year}{1958}).

\bibitem{Guta2009}
\bibinfo{author}{{Kahn}, J.} \& \bibinfo{author}{{Gu{\c t}{\u a}}, M.}
\newblock \bibinfo{title}{{Local Asymptotic Normality for Finite Dimensional
  Quantum Systems}}.
\newblock \emph{\bibinfo{journal}{Commun. Math. Phys.}}
  \textbf{\bibinfo{volume}{289}}, \bibinfo{pages}{597--652}
  (\bibinfo{year}{2009}).

\bibitem{Giovannetti2004}
\bibinfo{author}{Giovannetti, V.}, \bibinfo{author}{Lloyd, S.} \&
  \bibinfo{author}{Maccone, L.}
\newblock \bibinfo{title}{Quantum-enhanced measurements: beating the standard
  quantum limit}.
\newblock \emph{\bibinfo{journal}{Science}} \textbf{\bibinfo{volume}{306}},
  \bibinfo{pages}{1330--1336} (\bibinfo{year}{2004}).

\bibitem{Giovannetti2006}
\bibinfo{author}{Giovannetti, V.}, \bibinfo{author}{Lloyd, S.} \&
  \bibinfo{author}{Maccone, L.}
\newblock \bibinfo{title}{Quantum metrology}.
\newblock \emph{\bibinfo{journal}{Phys. Rev. Lett.}}
  \textbf{\bibinfo{volume}{96}}, \bibinfo{pages}{010401}
  (\bibinfo{year}{2006}).

\bibitem{Braunstein1994}
\bibinfo{author}{Braunstein, S.~L.} \& \bibinfo{author}{Caves, C.~M.}
\newblock \bibinfo{title}{Statistical distance and the geometry of quantum
  states}.
\newblock \emph{\bibinfo{journal}{Phys. Rev. Lett.}}
  \textbf{\bibinfo{volume}{72}}, \bibinfo{pages}{3439--3443}
  (\bibinfo{year}{1994}).

\bibitem{Berry2000}
\bibinfo{author}{Berry, D.~W.} \& \bibinfo{author}{Wiseman, H.~M.}
\newblock \bibinfo{title}{Optimal states and almost optimal adaptive
  measurements for quantum interferometry}.
\newblock \emph{\bibinfo{journal}{Phys. Rev. Lett.}}
  \textbf{\bibinfo{volume}{85}}, \bibinfo{pages}{5098--5101}
  (\bibinfo{year}{2000}).

\bibitem{Zwierz2010}
\bibinfo{author}{Zwierz, M.}, \bibinfo{author}{Pérez-Delgado, C.~A.} \&
  \bibinfo{author}{Kok, P.}
\newblock \bibinfo{title}{General optimality of the heisenberg limit for
  quantum metrology}.
\newblock \emph{\bibinfo{journal}{Phys. Rev. Lett.}}
  \textbf{\bibinfo{volume}{105}}, \bibinfo{pages}{180402}
  (\bibinfo{year}{2010}).

\bibitem{Mitchell2004}
\bibinfo{author}{Mitchell, M.~W.}, \bibinfo{author}{Lundeen, J.~S.} \&
  \bibinfo{author}{Steinberg, A.~M.}
\newblock \bibinfo{title}{Super-resolving phase measurements with a
  multi-photon entangled state}.
\newblock \emph{\bibinfo{journal}{Nature}} \textbf{\bibinfo{volume}{429}},
  \bibinfo{pages}{161--164} (\bibinfo{year}{2004}).

\bibitem{Eisenberg2005}
\bibinfo{author}{Eisenberg, H.~S.}, \bibinfo{author}{Hodelin, J.~F.},
  \bibinfo{author}{Khoury, G.} \& \bibinfo{author}{Bouwmeester, D.}
\newblock \bibinfo{title}{Multiphoton path entanglement by nonlocal bunching}.
\newblock \emph{\bibinfo{journal}{Phys. Rev. Lett.}}
  \textbf{\bibinfo{volume}{94}}, \bibinfo{pages}{090502}
  (\bibinfo{year}{2005}).

\bibitem{Nagata2007}
\bibinfo{author}{Nagata, T.}, \bibinfo{author}{Okamoto, R.},
  \bibinfo{author}{O'Brien, J.~L.}, \bibinfo{author}{Sasaki, K.} \&
  \bibinfo{author}{Takeuchi, S.}
\newblock \bibinfo{title}{Beating the standard quantum limit with four
  entangled photons}.
\newblock \emph{\bibinfo{journal}{Science}} \textbf{\bibinfo{volume}{316}},
  \bibinfo{pages}{726--729} (\bibinfo{year}{2007}).

\bibitem{Resch2007}
\bibinfo{author}{Resch, K.~J.} \emph{et~al.}
\newblock \bibinfo{title}{Time-reversal and super-resolving phase
  measurements}.
\newblock \emph{\bibinfo{journal}{Phys. Rev. Lett.}}
  \textbf{\bibinfo{volume}{98}}, \bibinfo{pages}{223601}
  (\bibinfo{year}{2007}).



\bibitem{Higgins2009}
\bibinfo{author}{Higgins, B.~L.} \emph{et~al.}
\newblock \bibinfo{title}{Demonstrating heisenberg-limited unambiguous phase
  estimation without adaptive measurements}.
\newblock \emph{\bibinfo{journal}{New J. Phys.}} \textbf{\bibinfo{volume}{11}},
  \bibinfo{pages}{073023} (\bibinfo{year}{2009}).

\bibitem{LIGO2011}
\bibinfo{author}{LIGO{\;}Collaboration}.
\newblock \bibinfo{title}{A gravitational wave observatory operating beyond the
  quantum shot-noise limit}.
\newblock \emph{\bibinfo{journal}{Nature Phys.}} \textbf{\bibinfo{volume}{7}},
  \bibinfo{pages}{962--965} (\bibinfo{year}{2011}).

\bibitem{Goda2008}
\bibinfo{author}{Goda, K.} \emph{et~al.}
\newblock \bibinfo{title}{A quantum-enhanced prototype gravitational-wave
  detector}.
\newblock \emph{\bibinfo{journal}{Nature Phys.}} \textbf{\bibinfo{volume}{4}},
  \bibinfo{pages}{472--476} (\bibinfo{year}{2008}).

\bibitem{Wineland1992}
\bibinfo{author}{D.~J.~Wineland, W. M.~I., J. J.~Bollinger} \&
  \bibinfo{author}{Moore, F.~L.}
\newblock \bibinfo{title}{Spin squeezing and reduced quantum noise in
  spectroscopy}.
\newblock \emph{\bibinfo{journal}{Phys. Rev. A}} \textbf{\bibinfo{volume}{46}},
  \bibinfo{pages}{R6797--R6800} (\bibinfo{year}{1992}).

\bibitem{Huelga1997}
\bibinfo{author}{Huelga, S.~F.} \emph{et~al.}
\newblock \bibinfo{title}{Improvement of frequency standards with quantum
  entanglement}.
\newblock \emph{\bibinfo{journal}{Phys. Rev. Lett.}}
  \textbf{\bibinfo{volume}{79}}, \bibinfo{pages}{3865--3868}
  (\bibinfo{year}{1997}).

\bibitem{Meyer2001}
\bibinfo{author}{Meyer, V.} \emph{et~al.}
\newblock \bibinfo{title}{Experimental demonstration of entanglement-enhanced
  rotation angle estimation using trapped ions}.
\newblock \emph{\bibinfo{journal}{Phys. Rev. Lett.}}
  \textbf{\bibinfo{volume}{86}}, \bibinfo{pages}{5870--5873}
  (\bibinfo{year}{2001}).

\bibitem{Leibfried2004}
\bibinfo{author}{Leibfried, D.} \emph{et~al.}
\newblock \bibinfo{title}{Toward heisenberg-limited spectroscopy with
  multiparticle entangled states}.
\newblock \emph{\bibinfo{journal}{Science}} \textbf{\bibinfo{volume}{304}},
  \bibinfo{pages}{1476--1478} (\bibinfo{year}{2004}).

\bibitem{Wasilewski2010}
\bibinfo{author}{Wasilewski, W.} \emph{et~al.}
\newblock \bibinfo{title}{Quantum noise limited and entanglement-assisted
  magnetometry}.
\newblock \emph{\bibinfo{journal}{Phys. Rev. Lett.}}
  \textbf{\bibinfo{volume}{104}}, \bibinfo{pages}{133601}
  (\bibinfo{year}{2010}).

\bibitem{Koschorreck2011}
\bibinfo{author}{Koschorreck, M.}, \bibinfo{author}{Napolitano, M.},
  \bibinfo{author}{Dubost, B.} \& \bibinfo{author}{Mitchell, M.~W.}
\newblock \bibinfo{title}{Sub-projection-noise sensitivity in broadband atomic
  magnetometry}.
\newblock \emph{\bibinfo{journal}{Phys. Rev. Lett.}}
  \textbf{\bibinfo{volume}{104}}, \bibinfo{pages}{093602}
  (\bibinfo{year}{2010}).


\bibitem{Higgins2007}
\bibinfo{author}{Higgins, B.~L.}, \bibinfo{author}{Berry, D.~W.},
  \bibinfo{author}{Bartlett, S.~D.}, \bibinfo{author}{Wiseman, H.~M.} \&
  \bibinfo{author}{Pryde, G.~J.}
\newblock \bibinfo{title}{Entanglement-free heisenberg-limited phase
  estimation}.
\newblock \emph{\bibinfo{journal}{Nature}} \textbf{\bibinfo{volume}{450}},
  \bibinfo{pages}{393--396} (\bibinfo{year}{2007}).

\bibitem{Demkowicz2010}
\bibinfo{author}{Demkowicz-Dobrzanski, R},
\newblock \bibinfo{title}{Multi-pass classical vs. quantum strategies in lossy phase estimation}.
\newblock \emph{\bibinfo{journal}{Laser Phys.}} \textbf{\bibinfo{volume}{20}},
  \bibinfo{pages}{1197-1202} (\bibinfo{year}{2010}).

\bibitem{Boixo2007}
\bibinfo{author}{Boixo, S}, \bibinfo{author}{Flammia, S.~T.},
  \bibinfo{author}{Caves, S.~T.}  \& \bibinfo{author}{Geremia, J.~M.}
\newblock \bibinfo{title}{Generalized Limits for Single-Parameter Quantum Estimation}.
\newblock \emph{\bibinfo{journal}{Phys. Rev. Lett}}
  \textbf{\bibinfo{volume}{98}}, \bibinfo{pages}{090401}
  (\bibinfo{year}{2007}).


\bibitem{Napolitano2011}
\bibinfo{author}{Napolitano, M.}, \bibinfo{author}{Koschorreck, M.},
  \bibinfo{author}{Dubost, B.}, \bibinfo{author}{Behbood, M.}, \bibinfo{author}{Sewell, R.~J.} \& \bibinfo{author}{Mitchell, M.~W.}
\newblock \bibinfo{title}{Interaction-based quantum metrology showing scaling beyond the Heisenberg limit}.
\newblock \emph{\bibinfo{journal}{Nature}}
  \textbf{\bibinfo{volume}{471}}, \bibinfo{pages}{486--489}
  (\bibinfo{year}{2011}).

\bibitem{Dorner2008}
\bibinfo{author}{Dorner, U.} \emph{et~al.}
\newblock \bibinfo{title}{Optimal quantum phase estimation}.
\newblock \emph{\bibinfo{journal}{Phys. Rev. Lett.}}
  \textbf{\bibinfo{volume}{102}}, \bibinfo{pages}{040403}
  (\bibinfo{year}{2009}).

\bibitem{Demkowicz2009a}
\bibinfo{author}{Demkowicz-Dobrza{\'n}ski, R.} \emph{et~al.}
\newblock \bibinfo{title}{Quantum phase estimation with lossy interferometers}.
\newblock \emph{\bibinfo{journal}{Phys. Rev. A}} \textbf{\bibinfo{volume}{80}},
  \bibinfo{pages}{013825} (\bibinfo{year}{2009}).

\bibitem{Shaji2007}
\bibinfo{author}{Shaji, A.} \& \bibinfo{author}{Caves, C.~M.}
\newblock \bibinfo{title}{Qubit metrology and decoherence}.
\newblock \emph{\bibinfo{journal}{Phys. Rev. A}} \textbf{\bibinfo{volume}{76}},
  \bibinfo{pages}{032111} (\bibinfo{year}{2007}).

\bibitem{Andre2004}
\bibinfo{author}{Andr{\`e}, A.}, \bibinfo{author}{S{\/0}rensen, A.~S.} \&
  \bibinfo{author}{Lukin, M.~D.}
\newblock \bibinfo{title}{Stability of atomic clocks based on entangled atoms}.
\newblock \emph{\bibinfo{journal}{Phys. Rev. Lett.}}
  \textbf{\bibinfo{volume}{92}}, \bibinfo{pages}{230801}
  (\bibinfo{year}{2004}).

\bibitem{Meiser2008}
\bibinfo{author}{Meiser, D.} \& \bibinfo{author}{Holland, M.~J.}
\newblock \bibinfo{title}{Robustness of heisenberg-limited interferometry with
  balanced fock states}.
\newblock \emph{\bibinfo{journal}{New J. Phys.}} \textbf{\bibinfo{volume}{11}},
  \bibinfo{pages}{033002} (\bibinfo{year}{2009}).

\bibitem{Kacprowicz2010}
\bibinfo{author}{Kacprowicz, M.}, \bibinfo{author}{Demkowicz-Dobrzanski, R.},
  \bibinfo{author}{Wasilewski, W.}, \bibinfo{author}{Banaszek, K.} \&
  \bibinfo{author}{Walmsley, I.~A.}
\newblock \bibinfo{title}{Experimental quantum enhanced phase-estimation in the
  presence of loss}.
\newblock \emph{\bibinfo{journal}{Nature Photon.}}
  \textbf{\bibinfo{volume}{4}}, \bibinfo{pages}{357--360}
  (\bibinfo{year}{2010}).

\bibitem{Kolodynski2010}
\bibinfo{author}{Ko{\l}ody{\'n}ski, J.} \&
  \bibinfo{author}{Demkowicz-Dobrza{\'n}ski, R.}
\newblock \bibinfo{title}{Phase estimation without a priori phase knowledge in
  the presence of loss}.
\newblock \emph{\bibinfo{journal}{Phys, Rev. A}} \textbf{\bibinfo{volume}{82}},
  \bibinfo{pages}{053804} (\bibinfo{year}{2010}).

\bibitem{Knysh2011}
\bibinfo{author}{Knysh, S.}, \bibinfo{author}{Smelyanskiy, V.~N.} \&
  \bibinfo{author}{Durkin, G.~A.}
\newblock \bibinfo{title}{Scaling laws for precision in quantum interferometry
  and the bifurcation landscape of the optimal state}.
\newblock \emph{\bibinfo{journal}{Phys. Rev. A}} \textbf{\bibinfo{volume}{83}},
  \bibinfo{pages}{021804(R)} (\bibinfo{year}{2011}).

\bibitem{Escher2011}
\bibinfo{author}{Escher, B.~M.}, \bibinfo{author}{de~Matos~Filho, R.~L.} \&
  \bibinfo{author}{Davidovich, L.}
\newblock \bibinfo{title}{General framework for estimating the ultimate
  precision limit in noisy quantum-enhanced metrology}.
\newblock \emph{\bibinfo{journal}{Nature Phys.}} \textbf{\bibinfo{volume}{7}},
  \bibinfo{pages}{406--411} (\bibinfo{year}{2011}).

\bibitem{Matsumoto2010}
\bibinfo{author}{Matsumoto, K.}
\newblock \bibinfo{title}{On metric of quantum channel spaces}.
\newblock \emph{\bibinfo{journal}{arXiv}} \bibinfo{pages}{1006.0300v1}
  (\bibinfo{year}{2010}).

\bibitem{Fujiwara2008}
\bibinfo{author}{Fujiwara, A.} \& \bibinfo{author}{Imai, H.}
\newblock \bibinfo{title}{A fibre bundle over manifolds of quantum channels and
  its application to quantum statistics}.
\newblock \emph{\bibinfo{journal}{J. Phys. A: Math. Theor.}}
  \textbf{\bibinfo{volume}{41}}, \bibinfo{pages}{255304}
  (\bibinfo{year}{2008}).

\bibitem{Bollinger1996}
\bibinfo{author}{Bollinger, J. J.~.}, \bibinfo{author}{Itano, W.~M.},
  \bibinfo{author}{Wineland, D.~J.} \& \bibinfo{author}{Heinzen, D.~J.}
\newblock \bibinfo{title}{Optimal frequency measurements with maximally
  correlated states}.
\newblock \emph{\bibinfo{journal}{Phys. Rev. A}} \textbf{\bibinfo{volume}{54}},
  \bibinfo{pages}{R4649--R4652} (\bibinfo{year}{1996}).

\bibitem{Dowling1998}
\bibinfo{author}{Dowling, J.~P.}
\newblock \bibinfo{title}{Correlated input-port, matter-wave interferometer:
  Quantum-noise limits to the atom-laser gyroscope}.
\newblock \emph{\bibinfo{journal}{Phys, Rev. A}} \textbf{\bibinfo{volume}{57}},
  \bibinfo{pages}{4736--4746} (\bibinfo{year}{1998}).

\bibitem{Bengtsson2006}
\bibinfo{author}{Bengtsson, I.} \& \bibinfo{author}{Zyczkowski, K.}
\newblock \emph{\bibinfo{title}{Geometry of quantum states: an introduction to
  quantum entanglement}} (\bibinfo{publisher}{Cambridge Univeristy Press},
  \bibinfo{year}{2006}).

\bibitem{Kay1993}
\bibinfo{author}{Kay, S.~M.}
\newblock \emph{\bibinfo{title}{Fundamentals of Statistical Signal Processing:
  Estimation Theory}} (\bibinfo{publisher}{Prentice Hall},
  \bibinfo{year}{1993}).

\bibitem{Nielsen2000}
\bibinfo{author}{Nielsen, M.~A.} \& \bibinfo{author}{Chuang, I.~L.}
\newblock \emph{\bibinfo{title}{Quantum Computing and Quantum Information}}
  (\bibinfo{publisher}{Cambridge University Press}, \bibinfo{year}{2000}).

\bibitem{Ruskai2001}
\bibinfo{author}{{Ruskai}, M.~B.}, \bibinfo{author}{{Szarek}, S.} \&
  \bibinfo{author}{{Werner}, E.}
\newblock \bibinfo{title}{{An Analysis of Completely-Positive Trace-Preserving
  Maps on 2x2 Matrices}}.
\newblock \emph{\bibinfo{journal}{Lin. Alg. Appl.}}
  \textbf{\bibinfo{volume}{347}}, \bibinfo{pages}{159--187}
  (\bibinfo{year}{2002}).

\bibitem{Molmer1996}
\bibinfo{author}{Molmer, K.}
\newblock \bibinfo{title}{Optical coherence: A convenient fiction}.
\newblock \emph{\bibinfo{journal}{Phys. Rev. A}} \textbf{\bibinfo{volume}{55}},
  \bibinfo{pages}{3195} (\bibinfo{year}{1996}).

\bibitem{Jarzyna2012}
\bibinfo{author}{Jarzyna, M.} \& \bibinfo{author}{Demkowicz-Dobrzanski, R.},
\newblock \bibinfo{title}{Quantum interferometry with and without an external phase reference}.
\newblock \emph{\bibinfo{journal}{Phys. Rev. A}} \textbf{\bibinfo{volume}{85}},
  \bibinfo{pages}{011801(R)} (\bibinfo{year}{2012}).

\bibitem{Fujiwara2001}
\bibinfo{author}{Fujiwara, A.}
\newblock \bibinfo{title}{Quantum channel identification problem}.
\newblock \emph{\bibinfo{journal}{Phys. Rev. A}} \textbf{\bibinfo{volume}{63}},
  \bibinfo{pages}{042304} (\bibinfo{year}{2001}).

\bibitem{Jamiolkowski1972}
\bibinfo{author}{Jamiolkowski, A.}
\newblock \bibinfo{title}{Linear transformations which preserve trace and
  positive semidefiniteness of operators}.
\newblock \emph{\bibinfo{journal}{Rep. Math. Phys}}
  \textbf{\bibinfo{volume}{3}}, \bibinfo{pages}{275--278}
  (\bibinfo{year}{1972}).

\bibitem{Choi1975}
\bibinfo{author}{Choi, M.-D.}
\newblock \bibinfo{title}{Completely positive linear maps on complex matrices}.
\newblock \emph{\bibinfo{journal}{Lin. Algebr. Appl.}}
  \textbf{\bibinfo{volume}{10}}, \bibinfo{pages}{285--290}
  (\bibinfo{year}{1975}).

\bibitem{CVX}
\bibinfo{author}{Grant M.} \& \bibinfo{author}{Boyd. S}
\bibinfo{title}{CVX: Matlab software for disciplined convex programming}, \bibinfo{url}{http://cvxr.com/cvx/} (\bibinfo{year}{2011}).

\end{thebibliography}

\end{document}